 \documentclass[final,5p,times,twocolumn]{elsarticle}

\usepackage{graphicx}

\usepackage{amssymb}

\usepackage{ulem} 
\usepackage[usenames]{color}
\newcommand{\be}{\begin{eqnarray}}
\newcommand{\ee}{\end{eqnarray}}
\newcommand{\non}{\nonumber \\}
\newcommand{\po}{{\rm P}}
\newcommand{\npo}{{\rm NP}}
\newcommand{\resi}{a_{-1}}

\begin{document}

\begin{frontmatter}


\author[juli]{M.~D\"oring}
\ead{m.doering@fz-juelich.de}
\author[juli,ias]{C.~Hanhart}
\author[uga]{F.~Huang}
\author[juli,ias]{S.~Krewald} 
\author[juli,ias,bonn]{U.-G.~Mei\ss ner} 
\address[juli]{Institut f\"ur Kernphysik and J\"ulich Center for Hadron Physics, 
Forschungszentrum J\"ulich, D-52425 J\"ulich,Germany}
\address[ias]{Institute for Advanced Simulation,
Forschungszentrum J\"ulich, D-52425 J\"ulich,Germany}
\address[uga]{Department of Physics and Astronomy, University of Georgia, Athens, GA 30602, USA}
\address[bonn]{Helmholtz-Institut f\"ur Strahlen- und Kernphysik (Theorie) and Bethe Center for Theoretical Physics, 
Universit\"at Bonn, Nu\ss allee 14-16, D-53115 Bonn, Germany
}
\title{\hspace{14cm}{\tiny FZJ-IKP-TH-2009-8, HISKP-TH-09/08}
\\
The role of the background in the extraction of resonance contributions from meson-baryon scattering}




\begin{abstract}
The separation of background and resonance contributions in pion-nucleon
scattering is an often discussed issue. We investigate
to what extent the background can be separated from the pole
contribution. For illustration we use results from an analytic model for
the meson-baryon interaction derived from meson exchange. We focus on the two
distinct cases of an elastic and  a highly inelastic resonance, namely
the $\Delta(1232)$ and the $\Delta^*(1700)$. Our results are also relevant for 
studies of dynamically generated resonances and attempts to extract bare
quantities from hadronic models to be compared to quark model results.
\end{abstract}


\end{frontmatter}



%
%
%
%
%
%
%
%
%
%
%
%

\section{Introduction}
The $\pi N$ interaction is one of the main sources of information
about the baryon spectrum, which is presently under experimental investigation, see e.g. Ref. \cite{nstar07}.
Information about the mass, width, and decay of baryon resonances
serves as a testing ground for models of the internal structures of
the nucleon and its excited states.

Most of the four and three star resonances listed by the PDG \cite{Amsler:2008zzb} have been obtained by
partial wave analyses followed by a model dependent analysis of the partial
wave amplitudes e.g. in terms of a background and Breit Wigner resonances 
\cite{Arndt:2006bf,Cutkosky:1979fy}. 
In the energy range between 2 and 3 GeV, presently under experimental
investigation, 
resonances start to overlap and the background may show some non-trivial structures. This situation calls for more sophisticated theoretical analyses. E.g., in the
partial wave analyses of Refs. \cite{Arndt:2006bf,Workman:2008iv}, poles in the
complex plane of the scattering energy are determined that are
identified with resonances. 
Models of the $K$ matrix type
\cite{Sarantsev:2005tg,Anisovich:2005tf,Manley:1992yb,Feuster:1997pq,Shklyar:2004ba,Vrana:1999nt}
and unitarized meson exchange models
\cite{Sato:1996gk,Surya:1995ur,Schutz:1998jx,Krehl:1999km,Gasparyan:2003fp} have
been constructed in the past to access pion nucleon scattering.

Another issue of relevance in this context is 
the question, if it is possible to remove
the hadronic contributions from observables in
a model independent way to allow access to 
quantities that can be identified with those
calculated from the quark model~\cite{leenew,svarcundress} --- see
also Ref.~\cite{nstar} for a recent discussion
of the subject. Such an analysis assumes
that a clean cut separation of pole and non-pole 
parts is possible. Also this issue will be discussed below.

And last but not least there is an increasing number of publications
claiming a molecular nature of various hadrons mainly based on
amplitudes from chiral perturbation theory unitarized by some means.
For the case of the $N^*(1535)$ considered later, see e.g.
Refs.~\cite{Kaiser:1995cy,Inoue:2001ip,Nieves:2001wt}. Thus, those singularities of the
$S$--matrix emerge from the iteration of background terms and not via
the inclusion of $s$--channel poles.  In most cases in these analyses
there was typically only a qualitative agreement of the theoretical
results with experimental data. In this work we will demonstrate that
such a procedure might be misleading: if an inclusion of a genuine
pole term is necessary to achieve a high quality description of the
data, this pole might well repel strongly the dynamically generated
one and lead to a very different picture.  The importance of a high quality description of the data was already stressed in Ref.~\cite{Meissner:1999vr}, where $\pi N$ scattering
was analyzed within the chiral unitary approach and the
interplay of genuine and dynamically generated resonances was
thoroughly investigated.  In any case, observable quantities that
allow one to distinguish between hadronic molecules and more
elementary states are urgently called for. For $s$--wave states close
to thresholds this is discussed in Refs.~\cite{evidence}.


\section{Generalities}

Within a theoretical model it is always possible to
 separate an amplitude into a pole and a non-pole part 
\be
T=T^\po+T^\npo
\label{deco1}
\ee where the pole part $T^\po$ is defined as the set of diagrams that
is 1-particle reducible, i.e. there is at least one $s$-channel
exchange. Usually, the non-pole, 1-particle irreducible part $T^\npo$
comes from meson exchange and $u$-channel processes collected into the
non-pole potential $V^{\npo}$ which is then unitarized using some
dynamical equation --- see Eq.~(\ref{dressed}) below. This
contribution is referred to as ``background''. The separation of the
type of Eq.~(\ref{deco1}) is widely used in the literature, see e.g.
\cite{Afnan:1980hp,Matsuyama:2006rp}. 

For a comparison with experiment, the poles and residues of the S-matrix 
are the relevant quantities. Therefore, we investigate
the Laurent
expansion around resonance poles. For an amplitude with a single
resonance and a pole at $z_0$ one may therefore write 
\be
T=\frac{\resi}{z-z_0}+a_0+{\cal O}(z-z_0) \ ,
\label{deco2}
\ee 
where $z$ is the scattering energy, $\resi$ is the residue, and
$a_0$ is a constant. 
In this study, we address the relation between $a_0$ and $T^\npo$.

To illustrate the discussion we use the amplitudes of the J\"ulich model,
an analytic coupled channel model based on meson exchange that respects two-body unitarity.  This model has been
developed over the past few years
\cite{Schutz:1998jx,Krehl:1999km}, with
its current form, as used in this study, given in Ref.
\cite{Gasparyan:2003fp}.
The coupled channel scattering equation is given by
\be
T=V+VGT
\label{bse}
\ee 
where indices and sums over intermediate quantum numbers have been
suppressed. The $VGT$ term implies an integration of the
three-momentum. $G$ is the intermediate meson baryon propagator of the
stable channels $\pi N$ and $\eta N$, and the channels involving
quasiparticles, $\sigma N$, $\rho N$, and $\pi\Delta$. The
pseudopotential $V$ iterated in Eq. (\ref{bse}) is constructed from an
effective interaction based on the Lagrangians of Wess and Zumino
\cite{Wess:1967jq}, supplemented by additional terms
\cite{Krehl:1999km,Gasparyan:2003fp} for including the $\Delta$ isobar, the
$\omega$, $\eta$, $a_0$ meson, and the $\sigma$. All these terms
contribute to the non-pole part. The pole part is given by baryonic
resonances up to $J=3/2$ that have been included in $V$ as bare $s$
channel propagators. The resonances obtain their width from the
rescattering provided by Eq. (\ref{bse}).

\begin{figure}
\begin{center}
\includegraphics[width=0.4\textwidth]{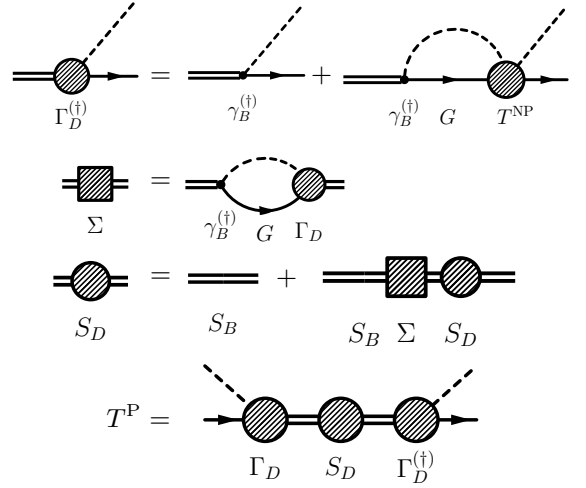}
\end{center}
\caption{Diagrammatic representation the bare vertices $\gamma_B$, dressed vertices $\Gamma_D$, self-energy $\Sigma$, dressed propagator $S_D$, and pole part $T^\po$.}
\label{fig:reorderhadro}
\end{figure}

In order to discuss the decomposition from Eq. (\ref{deco1}), it is necessary to determine the pole contribution $T^\po$ from the non-pole part $T^\npo$, i.e. from the set of diagrams that is 1-particle irreducible. For this, we define the quantities
\be
T^\npo(d,c)&=& V^{\npo}(d,c)+V^{\npo}(d,e)G(e)T^\npo(e,c) \non
\Gamma_D^{(\dagger)}(i,c)&=&\gamma_B^{(\dagger)}(i,c)+\gamma_B^{(\dagger)}(i,d)\,G(d)\,T^\npo(d,c)\non
\Gamma_D(c,i)&=&\gamma_B(c,i)+T^\npo(c,d)\,G(d)\,\gamma_B(d,i)\non
\Sigma(i,j)&=&\gamma_B^{(\dagger)}(i,c)\,G(c)\,\Gamma_D(c,j)
\label{dressed}
\ee 
where $\Gamma_D^{(\dagger)}$ ($\Gamma_D$) are the dressed creation
(annihilation) vertices and $\Sigma$ is the self-energy. Integrals and sums over intermediate states are not explicitly denoted in Eq. (\ref{dressed}). The bare
vertices $\gamma_B$ are derived from Lagrangians and provide bare
parameters that are fitted to the partial waves
\cite{Gasparyan:2003fp}. 
Note that for a simple energy and momentum independent $s$ wave interaction, $\gamma_B^{(\dagger)}=\gamma_B$ while for higher spin and partial waves the connection between bare creation and annihilation vertices can be more complicated.
The indices $i,j$ indicate the resonance,
while $c,d$ are indices in channel space. 
The dressed vertex, the self-energy and the dressed propagator are
schematically displayed in Fig.  \ref{fig:reorderhadro}.
Note, while for the stable channels ($\pi N$, $\eta N$), $T^\npo$ is individually two-body unitary, since it follows from solving a Lippmann Schwinger equation and $V^\npo$ is hermitian, $T^\po$ is not~\footnote{It should be mentioned that in principle the formalism is also three-body unitary for it follows closely that of Ref. \cite{Aaron:1969my}; however, the three-body unitarity is only approximate here, due to approximations in the $\sigma N$, $\rho N$, and $\pi\Delta$ propagators. But these technical details are irrelevant for the discussions of this paper.}.

With the quantities from Eq. (\ref{dressed}), the pole part is given by
\be
T^\po(c,c')&=&\Gamma_D(c,i)\,S_D(i,j)\,\Gamma_D^{(\dagger)}(j,c'),\non
S_D^{-1}&=&S_B^{-1}-\Sigma,\non
S_B^{-1}&=&z-M_0
\label{tpole}
\ee where $S_D$ ($S_B$) is the dressed (bare) resonance propagator and
$M_0$ is the bare mass. The pole part is indicated schematically in
Fig. \ref{fig:reorderhadro}.

One can expand the amplitude in a Laurent
series as shown in Eq. (\ref{deco2}). In fact, it is possible to
express $\resi$ and $a_0$ in terms of the dressed quantities from Eq.
(\ref{dressed}), \be
a_{-1}&=&\frac{\Gamma_D\,\Gamma_D^{(\dagger)}}{1-\frac{\partial}{\partial
    z}\Sigma}\non a_0&=&T^\npo+a_0^\po\non
a_0^\po&=&\frac{a_{-1}}{\Gamma_D\,\Gamma_D^{(\dagger)}}\,\left(
  \frac{\partial}{\partial
    z}(\Gamma_D\,\Gamma_D^{(\dagger)})+\frac{a_{-1}}{2}\,\frac{\partial^2}{\partial
    z^2}\,\Sigma\right).
\label{anala}
\ee All quantities on the right-hand side are evaluated at the pole
position $z=z_0$. 
As Eq. (\ref{anala}) shows, the constant $a_0$
receives a contribution from $T^\po$, which makes the identification
of $T^\npo$ as background problematic. This will be discussed in
detail in the next section.

For the pole search and the calculation of $\resi,\,a_0$, one has to
analytically continue the amplitude to unphysical sheets \cite{Suzuki:2008rp}. There are
several technical details to consider which are explained in
detail in Ref. \cite{Doring:2009yv}. With respect to the stable channels
$\pi N$ and $\eta N$, the poles are searched for on the second sheet if the pole
position is above threshold.  If the pole is below threshold it is on
the first or the second sheet for a bound and a virtual state,
respectively.  With respect to the unstable channels $\sigma N$, $\rho
N$, and $\pi\Delta$, the poles are searched for on the sheet that is obtained from
the analytic continuation of the self-energy of the unstable particle;
poles on the third and fourth $\sigma N$, $\rho N$, or $\pi\Delta$
sheets contribute little to the amplitude on the physical axis
\cite{Doring:2009yv}.

\section{Results}
\label{sec:resu}
\begin{figure}
\includegraphics[width=0.48\textwidth]{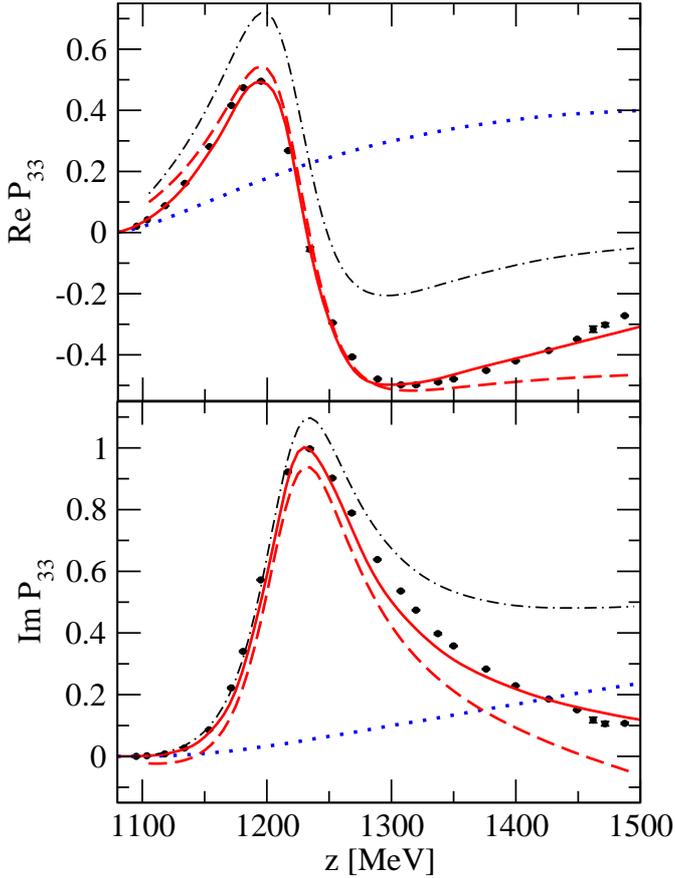}
\caption{The $P_{33}$ partial wave in $\pi N\to \pi N$. The data
  points are from the single energy solution of Ref. \cite{Arndt:2006bf}. The
  red solid (blue dotted) lines show the full $T$ (non-pole part $T^\npo$)
  from Eq. (\ref{deco1}). The black dashed dotted [red dashed] lines
  show the results from Eq. (\ref{appro1}) [Eq.
  (\ref{appro2})].}
\label{fig:p33tau}
\end{figure}

In Fig. \ref{fig:p33tau} the amplitude $\tau$ for the
$P_{33}$ partial wave is plotted. The connection to the $T$ matrix from Eq. (\ref{deco1}) is given by
\be
\tau=-\frac{\pi\,k\,E\,\omega}{z}\,T
\ee
where $k(E,\omega)$ are the on-shell three momentum (nucleon, pion energies) of the $\pi N$ channel.
The data points in Fig. \ref{fig:p33tau} refer to the single energy solution (SES) of the GWDAC
partial wave analysis \cite{Arndt:2006bf} for $\pi N\to \pi N$. The
red solid lines show the full solution of the J\"ulich model in its
current form \cite{Gasparyan:2003fp}, i.e. $T$ from Eq. (\ref{deco1}).
The blue dotted lines show the non-pole part $T^\npo$ as defined in the
decomposition from Eq. (\ref{deco1}).

As we have already seen in the discussion of Eq. (\ref{anala}), it is
not clear what should be considered as ``background'' of the
amplitude. 
To understand whether the decomposition from Eq. (\ref{deco1}) has a physical meaning, we make the following tests: For a given residue, extracted from the amplitude of the J\"ulich model, we add $T^\npo$,
\be 
T_1(z)\sim a_{-1}/(z-z_0)+T^{NP}(z).
\label{appro1}
\ee 
The results are shown in Fig. \ref{fig:p33tau} as the black dashed dotted line.  
As an alternative, one may choose to parameterize the
data via a Laurent series around the pole. This 
gives
\be T_2(z)\sim
a_{-1}/(z-z_0)+a_0
\label{appro2}
\ee
and is shown as the red dashed lines in Fig. \ref{fig:p33tau}. 

For $P_{33}$, it is obvious that the full solution is better described
by Eq.~(\ref{appro2}) than by  Eq.~(\ref{appro1}). The reason is that there
is a large contribution $a_0^\po$ from the pole part $T^\po$ to $a_0$
according to Eq. (\ref{anala}).  Eq.~(\ref{appro1}) corresponds to
neglecting $a_0^\po$. Thus, identifying $T^\npo$ with
the background is not appropriate in this case; instead, a systematic expansion
around the pole position in a Laurent series takes the constant
contributions from both pole and non-pole part into account properly.

Comparing the results from Eqs. (\ref{appro1}) and  (\ref{appro2})
in Fig. \ref{fig:p33tau}, a partial cancellation becomes visible, i.e.
$a_0^\po\sim -T^\npo$. For another strong $\pi N$ resonance (not shown
here), the $N^*(1520)D_{13}$, we observe a
similar behavior, as shown in Table \ref{tab:azeros}.
\begin{table}
  \caption{The constant terms $a_0=T^\npo+a_0^\po$ ($\pi N\to\pi N$ channel) from the Laurent expansion around the pole positions, for some resonances, in units of $[10^{-7}\,\rm{MeV}^{-2}]$. The ratio $|(T^\npo+a_0^{\rm P})/T^\npo|$ is shown in the last column.}
\begin{center}
\begin{tabular}{llll}
 \hline\hline
\hspace*{1.5cm}			&$T^\npo$\hspace*{1.2cm}	&$a_0^\po$ \hspace*{1.2cm}	&Ratio		       \\
$\Delta\,\,\,(1232)\,P_{33}$	&$-16.7-3.57i$			&$17.1+10.6i$			&$0.4$		       \\
$N^*(1520)\,D_{13}$ 		&$-4.62-0.56i$			&$3.03+1.23i$			&$0.4$		       \\
$\Delta^*\,(1620)\,S_{31}$ 	&$9.01-6.37i$     		&$-1.21+0.24i$ 			&$0.9$		       \\
$\Delta^*\,(1700)\,D_{33}$	&$0.80-0.52i$     		&$0.40+0.11i$			&$1.3$		       \\
$N^*(1720)\,P_{13}$ 		&$1.76-0.10i$     		&$0.45-0.56i$			&$1.3$		       \\
$\Delta^*\,(1910)\,P_{31}$ 	&$4.58-2.76i$     		&$-0.78+0.24$ 			&$0.9$		       \\
\hline\hline
\end{tabular}
\end{center}
\label{tab:azeros}
\end{table}
In the last column, a ratio close to zero indicates the partial cancellation.

\begin{figure}
\begin{center}
\includegraphics[width=0.45\textwidth]{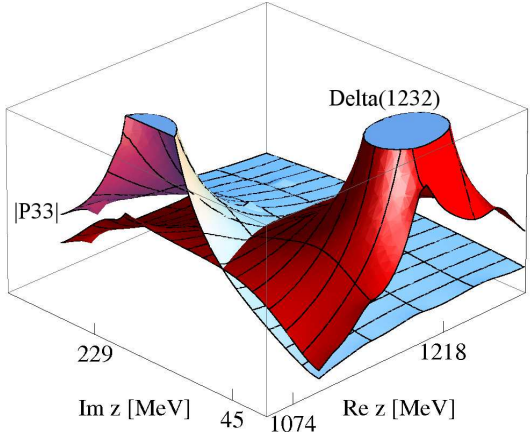}
\end{center}
\caption{The $P_{33}$ amplitude [arbitrary units] on the unphysical
  sheet as a function of $z$. The light blue surface shows
  $T^\npo$, the dark red surface $T=T^\po+T^\npo$ with the $\Delta(1232)$.
  Note that in the sum $T^\po+T^\npo$, the dynamically generated
  background pole at $\tilde{z}_0=1074+229\,i$ MeV has disappeared.}
\label{fig:bgresp33}
\end{figure}

In order to understand the underlying cancellation mechanism, the
complex $z$ plane in $P_{33}$ is inspected as shown in Fig.
\ref{fig:bgresp33}.
The light blue surface shows the non-pole part $T^{NP}$, the dark red
surface shows the full amplitude $T=T^P+T^{NP}$. In the full solution,
the large $\Delta(1232)$ pole at $z_0=1218+45\,i$ MeV is clearly
visible. Surprisingly, there is a pole in $T^{NP}$, close to the $\pi
N$ threshold and far in the complex plane at $\tilde{z}_0=1074+229\,i$ MeV.
This pole is dynamically generated from the unitarization provided by
the scattering equation (\ref{bse}). It is a nonperturbative structure
which on the physical axis appears as a substantial background (blue
dotted lines in Fig. \ref{fig:p33tau}). 

The attraction and subsequent pole formation in the unitarization could be traced back to the nucleon exchange potential in the $\pi N\to\pi N$ transition of the J\"ulich model (cf. Eq. (A5) of Ref. \cite{Krehl:1999km}). Switching off all other transitions and channels in $V^\npo$ of Eq. (\ref{dressed}), around half of the rise of Re $T^\npo$ in Fig. \ref{fig:p33tau} is obtained. Together with the correlated two pion exchange (cf. Eqs. (A6, A7) of Ref. \cite{Krehl:1999km}), large part of the rise can be explained. This result is confirmed by the fact that the $\pi N$ residue of the pole at $\tilde{z}_0=1074+229\,i$ MeV is $\sqrt{a_{-1}}=(11-18\,i)\cdot 10^{-3}$ MeV$^{-1/2}$, which is much larger than the residue to the other channels $\rho N$ and $\pi\Delta$.

Once the pole part $T^\po$ is
added to $T^\npo$, this pole has disappeared, i.e.
$T$ around $\tilde{z}_0$ shows no pole structure in Fig. \ref{fig:bgresp33}.

Poles in $T^\npo$ are indeed systematically cancelled in the sum
$T^\npo+T^\po$; the reason is the appearance of $T^\npo$ in $T^\po$ as
shown in Eq. (\ref{dressed}).  For a qualitative understanding, we
consider a one-channel case with separable potentials and a loop
function G. Given a bare coupling $b$, we can write the pole part
$T^\po$ matrix from Eqs. (\ref{dressed},\ref{tpole}) as 
\be
T^P&=&\frac{b^2\left(1+G\,T^\npo\right)^2}{z-M_0-\Sigma},\quad \Sigma
=b^2\,G\left(1+T^\npo G\right) \ .
\label{tpolesimple}
\ee 
The dynamically generated pole in the background of $P_{33}$ at
$z=\tilde{z}_0$ is to leading order given by the residue term with $\tilde{a}_{-1}$;
thus we approximate $T^\npo\sim \tilde{a}_{-1}/(z-\tilde{z}_0)$. Inserting this
expression in Eq. (\ref{tpolesimple}), we obtain 
\be
T^\npo+T^\po{=}\frac{b^2(z-\tilde{z}_0+\tilde{a}_{-1}\,G)+\tilde{a}_{-1}(z-M_0)}{(z-\tilde{z}_0)(z-M_0)-b^2\,G(z-\tilde{z}_0+\tilde{a}_{-1}\,G)}\non
\label{sumtptnp}
\ee 
which is finite at $z=\tilde{z}_0$ for $b\neq 0$, i.e. the pole at $\tilde{z}_0$
has disappeared. The denominator of Eq. (\ref{sumtptnp}) still has two
zeros, one at the physical $\Delta(1232)$ position, and another one
far in the complex plane (${\rm Im}\,z>400$ MeV) with only a small
contribution on the physical axis. The full amplitude on the real $z$
axis is dominated by the $\Delta(1232)$.

Furthermore, even in this qualitative model, it is possible to see
that for $a_0$, evaluated using Eq. (\ref{anala}), $a_0^\po\sim
-T^\npo$ at the pole position of the $\Delta(1232)$. This comes from
the $z$ dependence of the leading order term $T^\npo\sim
\tilde{a}_{-1}/(z-\tilde{z}_0)$, that enters in the calculation of $a_0^\po$ in Eq.
(\ref{anala}).

The picture is now complete: $T^\npo$ is a nonperturbative structure
associated with a dynamically generated pole in the complex plane.
Once the pole part with the $\Delta(1232)$ is added, the dynamically
generated pole is driven far into the complex plane. For the Laurent
expansion at the $\Delta(1232)$ pole, one finds
$a_0^\po\sim -T^\npo$; for the amplitude on the physical axis, this
corresponds to the cancellation of $T^\npo$, when the pole part is
added.  The full amplitude $T=T^\po+T^\npo$ is then dominated by a
clean $\Delta(1232)$ resonance.

This picture supports also the framework of isobar models, in which
the interaction is dominated by resonances alone: as we have seen,
large, nonperturbative structures in $T^\npo$ can be systematically
cancelled, resulting in a resonance dominated amplitude.

The cancellation of the pole as discussed following Eq. (\ref{sumtptnp}) always takes place. However, the poles can be weakly correlated, if, e.g., they are far away from each other. In such cases, although the pole is cancelled, it will reappear close to the original position and with similar residue, once the genuine pole is added. In this scenario, the genuine pole appears as a rather weak perturbation. This behavior has been confirmed in numerical simulations.

Another question concerns the situation, when the two interacting poles are on different Riemann sheets. Recently, an example of such a situation has been found (Sec. 3 of Ref.~\cite{Doring:2009yv}): within the J\"ulich model, a dynamically generated pole is found in the $D_{13}$ partial wave on the third (hidden) $\rho N$ sheet. This pole in $T^\npo$ at $\tilde{z}_0=1613-83\,i$ MeV is visible at the physical axis at around $z\sim 1.7$ GeV, because of the $\rho N$ branch point in the complex plane. The resonance is mainly generated from the attraction in the $\rho N$ channel; note a similar structure has been found recently in Ref. \cite{Oset:2009vf}.

However, there is another resonance in the same partial wave: Once the $N^*(1520)$ is added as a genuine state $(T^\po)$, it develops a pole on the second (non-hidden) $\rho N$ sheet. At the same time, the dynamically generated pole in $T^\npo$ on the third $\rho N$ sheet is driven far away to the complex plane in the sum $T=T^\po+T^\npo$. In $T$ on the physical axis, the resonant structure at around $z\sim 1.7$ GeV has disappeared, and only the dominant $N^*(1520)$ is visible in $\pi N$ scattering. We thus do not identify the dynamically generated pole with an $N^*(1700)D_{13}$ resonance~\cite{Amsler:2008zzb}.

In the $D_{13}$ partial wave, there is thus a similar mechanism of pole repulsion as previously discussed for $P_{33}$; indeed, the ratio in Table \ref{tab:azeros} is as small as for the $P_{33}$ partial wave. For the other partial waves listed in Table \ref{tab:azeros}, the ratio is much larger and indeed we could not find any poles in $T^\npo$ for these cases within the allowed range of Im $z<200$ MeV~\cite{Doring:2009yv}. 

While the extreme situation $a_0^P\simeq -T^\npo$ is tied to appearance of poles in $T^\npo$ within the present model, even for those cases with ratio closer to 1, $a_0^P$ is never zero; the pole part always contributes to the constant term, making an identification of $T^\npo$ as ``background'' questionable.

It should be clear that the presence of the pole in $T^{\npo}$ is a
property of the model and not a general one. E.g. $\pi N$ scattering
can well be studied within chiral perturbation theory with
an explicit $\Delta$ field~\cite{chptdelta,daniel}. Then, the background
amplitudes are treated perturbatively and accordingly no pole is
generated in the complex plane. On the other hand, pole terms may be
added in a consistent manner~\cite{daniel}. Correspondingly, the
contribution of the non-pole parts to the $\Delta(1232)$ properties
will be very different.  Thus, splitting the features of the $P_{33}$
partial wave into bare pole and background contribution is model
dependent and has no physical significance. Nonetheless, the physical pole of the $\Delta(1232)$ has been separated in the three analyses 
\cite{Arndt:2006bf,Cutkosky:1979fy,Hohler93} quoted by the PDG \cite{Amsler:2008zzb}, with pole
positions in close agreement. 

The splitting into $T^\po$ and $T^\npo$ is also responsible for the
renormalization of the bare vertex; indeed, the dressed vertex
$\Gamma_D$ depends on the bare vertex $\gamma_B$ and $T^\npo$ as Eq.
(\ref{dressed}) shows. This gives a negative answer to the question raised in the
Introduction, whether it is possible to extract bare quantities such
as $\gamma_B$ in a model independent way: In the J\"ulich model,
$|(\Gamma_D-\gamma_B)/\gamma_B|=0.45$ at the $\Delta(1232)$ pole
position, i.e. there is a 45 \% renormalization of the bare vertex,
coming from the model dependent part $T^\npo$. 

Finally, let us mention that within the analytic model  as given in Eq. (\ref{tpolesimple}) it
is easy to compensate a change in the bare coupling $b$ by a different regulator for the loop function $G$, with very similar phase
shifts on the physical axis. This further points at the model
dependence of the size of the bare coupling.

While the model dependence of $\gamma_B$ has been demonstrated, the
question remains whether the dressed vertex $\Gamma_D$ from Eq.
(\ref{dressed}) is a physically meaningful quantity. Yet, only the
residue provides a well defined expansion parameter of the amplitude~\cite{Djukanovic:2007bw}.
As Eq. (\ref{anala}) shows, there is a difference between $a_{-1}$
and $\Gamma_D$, given by the $Z$ factor $1-\Sigma'$. In the J\"ulich
model, for the $\Delta(1232)$ the ratio $|(g-\Gamma_D)/g|=0.40$
($g=\sqrt{a_{-1}}$) is also large which implies that $\Gamma_D$,
without the $Z$ factor, can substantially depend on the separation
into $T^\npo$ and $T^\po$.

\begin{figure}
\includegraphics[width=0.4\textwidth]{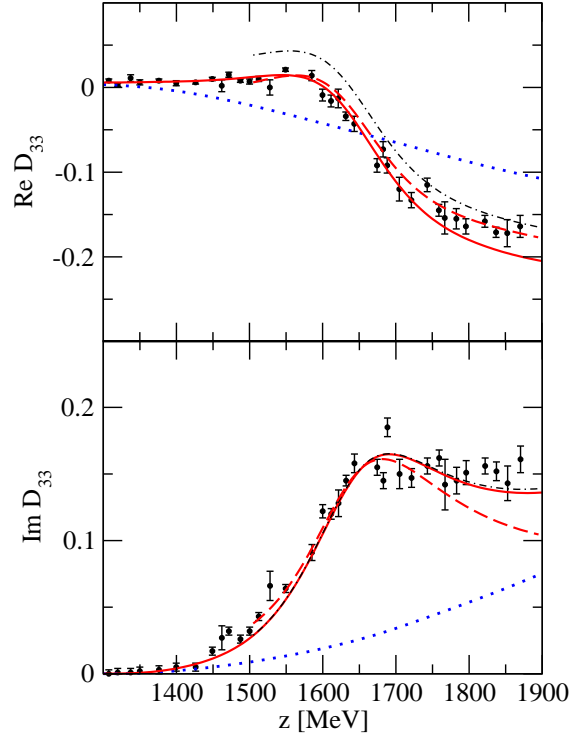}
\caption{The $D_{33}$ partial wave in $\pi N\to \pi N$. See Fig. \ref{fig:p33tau} for labels of the curves.}
\label{fig:d33tau}
\end{figure}

The observed cancellation $a_0^\po\sim - T^\npo$ is only given for the $\Delta(1232)$ and
$N^*(1520)$ as Table \ref{tab:azeros} shows. For other
resonances, the ratio in the last column is close to one, i.e.
$a_0^\po$ is small and there is no cancellation with $T^\npo$;
instead, the naive identification of $T^\npo$ as background is
justified reasonably well. As an example, we show the $D_{33}$
partial wave in Fig. \ref{fig:d33tau}.
The $\Delta(1700)D_{33}$ is a wide resonance with a pole at
$z_0=1637-118\,i$ MeV. It is quite
inelastic in $\pi N$ and there is a substantial background. In
principle, one could expect a cancellation similar to that of the
$\Delta(1232)$. This is, however, not the case and $a_0^\po$ is small
as Table \ref{tab:azeros} shows. Inspecting the complex $z$ plane of
the $D_{33}$ partial wave, there is no pole in $T^\npo$.

  The cancellation behavior $a_0^P\sim -T^\npo$ is only present for the low energy resonances as Table \ref{tab:azeros} shows. For the higher $N^*$'s and $\Delta^*$'s we inspected, the ratio is close to 1; even if there are dynamically generated resonances in $T^\npo$ at higher energies, their poles are so far in the complex plane that they hardly interact with the genuine states. In fact, for the partial waves of these higher states, we have found no poles in $T^\npo$ up to 
  $|{\rm Im}\,z|=400$ MeV.

In Table \ref{tab:azeros}, the ratio for the $\Delta^*(1700)$
  is 1.3, i.e. there is a certain but modest contribution coming from
  $a_0^\po$. In Fig. \ref{fig:d33tau} we show the background, full
  solution and the results from Eqs. (\ref{appro1},
  \ref{appro2}), with the same line styles as in Fig.
  \ref{fig:p33tau}. Similar as for the $P_{33}$, the systematic
  expansion in a Laurent series around the pole (dashed lines)
  delivers a better approximation in ${\rm Re\,D_{33}}$ 
than the naive
  summation of $T^\npo$ plus residue term (dashed dotted lines). Yet,
  above the pole position, the situation is opposite in ${\rm
    Im}\,D_{33}$, and Eq.~(\ref{appro1}) better. This is simply due to
  the fact that away from the pole position, higher order
  contributions in the Laurent expansion become important; while we
  consider the expansion only up to the $a_0$ term, $T^\npo$ includes
  those. Thus, the high energy tail of the
  $\Delta^*(1700)$ in ${\rm Im}\,D_{33}$ is better modelled by Eq.
  (\ref{appro1}) than by Eq. (\ref{appro2}).

  In any case, apart from these details, for the $D_{33}$ partial wave
  the naive picture to identify $T^\npo$ as background is qualitatively
  justified. As pointed out before, there is, other than in the
  $\Delta(1232)$ partial wave, no pole in $T^\npo$; although $T^\npo$
  is not small, it is perturbative.

The interaction of a pole in $T^\npo$ and the pole term $T^\po$,
as found here for the $\Delta(1232)$, has also implications for unitarized chiral
perturbation theory (U$\chi$PT), in which resonances appear dynamically generated
from unitarization \cite{Kaiser:1995cy,Inoue:2001ip,Nieves:2001wt}. We do not discuss the validity of such models here.
Yet, in some U$\chi$PT
models of the $N^*(1535)\,S_{11}$ \cite{Kaiser:1995cy,Inoue:2001ip}, the pole of
the $N^*(1650)\,S_{11}$ is not considered. In case of the $\Delta(1232)$, we have seen previously that the genuine pole term has a large impact both on position and properties of the dynamical resonance; for the $S_{11}$ this implies that the resonance interference of the two $N^*$'s
should not be neglected.

\section{Summary}
The standard decomposition into $T^\po$ and $T^\npo$ is model dependent. As far as the question of the background is concerned,
the pole part $T^\po$ provides a nonzero constant term $a_0^\po$ making the identification of $T^\npo$ as background problematic. 

The non-pole part $T^\npo$ from meson exchange and $u$-channel processes
can be large and provide a nonperturbative amplitude. This amplitude enters in the determination of bare and dressed vertices; thus, a physically meaningful measure of the resonance coupling strength, independent of the decomposition into pole and non-pole part, is only given by the residue.

In the $P_{33}$ $\pi N$ channel, $T^\npo$ is non-perturbative and has a pole in the complex plane.
However, such poles in $T^\npo$ can be systematically cancelled, once the
$s$-channel part $T^\po$ is added; as a consequence, the
full amplitude is dominated by a genuine resonance with a negligible background. Yet, as we have seen in
the $D_{33}$ partial wave, even a significant background can be perturbative, and
in this case the naive picture to consider $T^\npo$ as background is
justified.

\vspace*{0.2cm}

The work of M.D. is supported by DFG (Deutsche Forschungsgemeinschaft, GZ: DO 1302/1-1). This work is supported in part by the Helmholtz Association through funds provided to the virtual
institute ``Spin and Strong QCD'' (VH-VI-231), by the  EU-Research Infrastructure Integrating Activity
 ``Study of Strongly Interacting Matter" (HadronPhysics2, grant n. 227431)
under the Seventh Framework Program of EU and by the DFG (TR 16). F.H. is grateful to the support from the Alexander von Humboldt Foundation and the COSY FFE grant No. 41445282  (COSY-58).

\end{document}